# Octahedral coupling in (111)- and (001)-oriented La$_{2/3}$Sr$_{1/3}$MnO$_3$/SrTiO$_3$ heterostructures

*Magnus Moreau,[1] Sverre M. Selbach,[2] and Thomas Tybell[1,*]*
1) Department of Electronic Systems, NTNU Norwegian University of Science and Technology, 7491 Trondheim, Norway
2) Department of Materials Science and Engineering, NTNU Norwegian University of Science and Technology, 7491 Trondheim, Norway
*E-mail: thomas.tybell@iet.ntnu.no

**Abstract**
Rotations and distortions of oxygen octahedra in perovskites play a key role in determining their functional properties. Here we investigate how octahedral rotations can couple from one material to another in La$_{2/3}$Sr$_{1/3}$MnO$_3$/SrTiO$_3$ epitaxial heterostructures by first principles density functional theory (DFT) calculations, emphasizing the important differences between systems oriented perpendicular to the (111)- and (001)-facets. We find that the coupling length of out-of-phase octahedral rotations is independent of the crystalline facet, pointing towards a steric effect. However, the detailed octahedral structure across the interface is significantly different between the (111)- and (001)-orientations. For (001)-oriented interfaces, there is a clear difference whether the rotation axis in SrTiO$_3$ is parallel or perpendicular to the interface plane, while for the (111)-interface the different rotations axes in SrTiO$_3$ are symmetry equivalent. Finally, we show that octahedral coupling across the interface can be used to control the spatial distribution of the spin density.

**Introduction**

The strong structure property coupling in epitaxial perovskite oxide heterostructure can give rise to novel properties at the interface. Central to the properties of these materials are the oxygen octahedral rotations, which can template across an interface to induce properties not found in the bulk material [1-4]. Examples where octahedral coupling at (001)-oriented interfaces have induced different properties include: improper ferroelectricity in $PbTiO_3$/$SrTiO_3$ (STO) superlattices [5], increased magnetization and electrical conductivity at the interface between $La_{2/3}Sr_{1/3}MnO_3$ (LSMO) and $(LaAlO_3)_{0.3}(Sr_2AlTaO_6)_{0.7}$ [6], controlled magnetic anisotropy in LSMO on $NdGaO_3$ [7], reduced band gap in $BiFeO_3$ at the interface with LSMO grown on STO [8], and the ability to switch ferromagnetism on and off in $CaMnO_3$/$CaRuO_3$ superlattices [9]. To utilize these effects it is important know the length scale which the octahedral rotations couple across an interface. To this end, He *et al.* [10] used density functional theory (DFT) to calculate the octahedral coupling length from an infinitely rigid (001)-substrate without any chemical discrepancies, and found that the coupling length could be a short as one layer or extend deep into the film. This has later been confirmed experimentally by Aso *et al.* [11] who showed that the coupling length varied between one layer for $BaTiO_3$ on STO and seven layers for $Sr_{0.5}Ca_{0.5}TiO_3$ on STO.

A good model system to study the effect of octahedral coupling is ferromagnetic LSMO thin films grown on STO. Bulk STO has a tetragonal $I4/mcm$ ground state with $a^0a^0c^-$ Glazer tilt pattern [12], while above 105 K the structure is cubic with space group $Pm\overline{3}m$ and $a^0a^0a^0$ tilt pattern. Possible structural coupling due to octahedral rotations can hence be turned on and off experimentally by cooling below and heating above the transition temperature [13-18]. It has e.g. been found for (001)-oriented LSMO/STO that tetragonal STO can give rise to twinning in the LSMO thin film [13], alter the coercive field [14], transport properties [15], and magnetic anisotropy [16] of LSMO. Furthermore, Segal *et al.* [17] explained changes in LSMO transport properties around the STO transition temperature in terms of phonon coupling across the interface.

Recent experimental advances has enabled growth along other crystalline facets, such as the (111)-facet [19], a promising route to further develop oxide electronics [20]. The hexagonal symmetry, with a $S_6$ inversion axis, of the (111)-interface can allow for topologically protected states [21-26]. Furthermore, strain in the (111)-plane is considerably different compared to strain in the (001)-plane, and can result in structural Goldstone modes [27-29]. It has also been shown that octahedral rotation coupling can induce a net magnetic moment in antiferromagnetic $LaFeO_3$ without charge transfer [30]. However, a detailed understanding of how a (111)-interface affects the octahedral coupling is still missing. As shown in Figure 1, the oxygen octahedra couple through three oxygen at the octahedra face at the (111)-interface (Figure 1 a), while at the (001)-interface the octahedra couple through one apex oxygen (Figure 1 b). Thus, a different coupling is expected for the two facets. Furthermore, for the (111)-interface all of the three octahedra rotation axes $x, y$ and $z$ are equivalent. However, for the (001)-interface the $x$- and $y$- axes are parallel to the interface plane, while the $z$-axis is perpendicular to it. Hence, for the (111)-interface, no difference in the coupling between out-of-plane $\gamma$-rotations and in-plane $\alpha$- and $\beta$-rotations (see Figure 1 for definitions) is expected, in contrast to (001)-oriented interfaces [10].

In this work we compare octahedral coupling between LSMO thin films and STO substrates by DFT, with focus on the difference between (111)- and (001)-oriented interfaces. DFT has a long track record when it comes to evaluating octahedral rotations in perovskites, examples include how epitaxial strain affects octahedral rotations, and how octahedral rotations can couple across an interface [3, 10, 27-33]. We exploit the tilt patterns in the two stable phases of STO (tetragonal and cubic) to compare how octahedral rotations in the substrate can affect the film. The paper is structured as follows: First the computational details are explained. Then different methodologies for modelling the Sr doping in LSMO are compared, before establishing how strain from (111)- and (001)-oriented STO affects the octahedral rotations of LSMO. Finally, we correlate changes in octahedral rotations in LSMO with the spatial spin density distribution.

**Computational details**

The DFT calculations were performed with the Vienna Ab-initio Simulation Package (VASP, version 5.3.3) [34, 35] employing the projector augmented wave method (PAW) [35, 36]. The Perdew-Burke-Ernzerhof generalized gradient approximation for solids (PBEsol) was chosen as it has been shown to accurately reproduce the crystal structure and lattice parameters of solids [37]. The recommended PAW potentials supplied with VASP for La, Sr, Mn, Ti and O where used, having electron configurations $4s^2 4p^6 5d^1 6s^2$, $4s^2 4p^6 5s^2$, $3p^6 3d^5 4s^2$, $3s^2 3p^6 3d^2 4s^2$ and $2s^2 2p^4$, respectively. To treat the correlated $d$ and $f$ electrons of Mn and La, Hubbard U values of 3 and 10 eV, respectively, where applied to these orbitals using the Dudarev method [38]. These values U have been shown to adequately model perovskite oxides containing Mn and La [39-41]. To include the strain from the substrate, the in-plane lattice vectors were locked to those calculated for cubic SrTiO$_3$, while the out-of-plane lattice vector and atomic coordinates were allowed to relax. For (111)-strain, the in-plane lattice vectors were along $[1\bar{1}0]$- and $[01\bar{1}]$-pseudocubic directions and the out-of-plane lattice vector was along $[111]$-pseudocubic direction, while for (001)-strain the in-plane lattice vectors were along $[100]$- and $[010]$-pseudocubic directions and the out-of-plane lattice vector was along $[001]$-pseudocubic direction. The atomic positions and the free lattice parameters were relaxed until the forces on the atoms were below 1 meV/ Å, for the bulk and strain calculations. The plane wave cutoff energy was set to 550 eV, and the calculations of bulk properties were done with LSMO in the bulk $R\bar{3}c$ in hexagonal setting, which can be considered a $\sqrt{2} \times \sqrt{2} \times 2\sqrt{3}$ supercell of the aristotype perovskite structure. This cell is oriented such that the **a, b** and **c** lattice vectors are parallel to the $[1\bar{1}0]$-, $[01\bar{1}]$- and $[111]$-pseudocubic directions respectively, hence this cell was also used for calculations of LSMO when strained in the (111)-plane. For (001)-strain a $2 \times 2 \times 2$ supercell was used, with lattice vectors **a**, **b** and **c** along $[100]$-$[010]$- and $[001]$-psuedocubic directions, respectively. A $5 \times 5 \times 2$ gamma centered k-point mesh was used in the $\sqrt{2} \times \sqrt{2} \times 2\sqrt{3}$ calculation cells, while a $4 \times 4 \times 4$ mesh was used for the $2 \times 2 \times 2$ cells. Corresponding k-point densities were used for the supercells. To model Sr doping the virtual crystal approximation [42] (VCA) was used to create a superposition of the La and Sr PAW-potentials. As VCA has mainly been used to model LSMO with other DFT-codes previously [43], test calculations were performed. Bulk parameters of LSMO with VCA were compared to discrete doping with Sr distributed over the possible A sites in a $\sqrt{2} \times \sqrt{2} \times 2\sqrt{3}$ supercell containing 30 atoms, and a $2\sqrt{2} \times 2\sqrt{2} \times 2\sqrt{3}$ containing 120 atoms.

For the calculations of the interfaces, $\sqrt{2} \times \sqrt{2} \times 10\sqrt{3}$ and $\sqrt{2} \times \sqrt{2} \times 14$ supercells were used for the (111)- and (001)-interface, respectively, as illustrated in Figure 2. These cells were terminated with Ti for (111)-interfaces and TiO$_2$ for (001)-interfaces, as Ti-based terminations are the results of the commonly used substrate preparations for both STO(001) and (111) [44]. The (111)-interface calculation cells have 23.5 layers of LSMO and 6.5 layers of STO, with a layer spacing of $d_{111} \sim a_{pc}/\sqrt{3}$, where $a_{pc}$ is the pseudocubic lattice constant (Figure 2 a), while the (001)-interface calculation cells have 11.5 layers of LSMO and 2.5 layers of STO with a layer spacing of $d_{001} \sim a_{pc}$ (Figure 2 b). Octahedra in the STO layers away from the interface were fixed to have no rotations, equivalent to the cubic phase above 105 K, or fixed to have out-of-phase rotations around one of the pseudocubic axis, equivalent to the tetragonal phase stable below 105 K. Ion positions in LSMO and in the free STO layer were allowed to relax until the forces were less than 5 meV/Å, see Figure 2. For the (001)-interface, there are two symmetry equivalent ways to have the out-of-phase rotation axis of the tetragonal STO; it can be either out-of-plane (tilt pattern $a^0 a^0 c^-$) or in-plane (tilt pattern $a^0 b^- a^0$ equivalent to $a^- b^0 b^0$). While for the (111)-interface, the three different tetragonal STO tilt patterns, $a^0 a^0 c^-$, $a^0 b^- a^0$ and $a^- b^0 b^0$, are all symmetry equivalent. We note that octahedral rotations can couple more than two layers into an adjacent substrate [45, 46], however since the goal of this study is to investigate how the underlying STO layer effects the rotational pattern of LSMO we allow only the atoms of the two STO layers closest to the interface relax while the layers further away are locked to its bulk values, as illustrated in Figure 2. In order to isolate the effect of rotational coupling from that of the strain, the in-plane lattice parameters of the interface cells were locked to those calculated for cubic STO, while the out-of-plane lattice parameter was locked to the sum of STO and LSMO strained to STO with the given numbers of layers. It was also found that the small changes in lattice parameters between cubic and tetragonal STO has little influence on the octahedral rotations of LSMO, in agreement with Segal et al. [17]. All crystallographic directions are given in the pseudocubic setting unless otherwise specified, and the structure visualizations were done with VESTA [47].

**Results and discussion**

*LSMO modeled with VCA*

To test how suitable VCA is for modeling distortions in LSMO we compare results for bulk LSMO calculated with VCA to a supercell approach with different sizes, containing 30 and 120 atoms, in addition to experimental results [48]. As shown in Table I, the lattice parameters and the $c/\sqrt{6}a$-ratio are closest to the experimental value for the largest supercell, but still within a satisfactory range for the calculations by VCA. Importantly, the VCA and the supercell approaches all correctly reproduce the experimental magnetic moment. The average octahedral rotation angles are subtly underestimated in the supercells, while the VCA approach overestimates them. However, in the supercell approach, different Mn atoms have different distance to the Sr atoms, causing the octahedral rotation angles to vary for different octahedra in the supercell. This variation with position is due to the artificial ordering of Sr in the calculations. When studying octahedral coupling, such spread in octahedral rotations will be superimposed on any effects for octahedra coupling across the interface, making such approach less intuitive. To quantify this spread, the standard deviation in octahedral rotation, $\sigma$, is given in Table I. As shown, $\sigma$ is largest in the 120 atom cell and becomes smaller in the 30 atom cell, while it is exactly equal to zero for the VCA approach, as there is no artificial Sr ordering. This makes the VCA approach suitable when comparing octahedral coupling. Table I also shows how the calculated exchange energy depend on Sr doping model, albeit in all cases a ferromagnetic ground state is the most stable. Finally, we note that due to the increased symmetry and reduced cell size (compared to the large supercell) the VCA approach is less CPU expensive than the supercell approaches.

The calculated density of states (DOS) using the VCA and supercell approaches are shown in Figure 3. The DOSes are quite similar, especially around the Fermi level, and all approaches capture the known half-metallic character of LSMO, with a similar bandwidth of $\sim 2.95$ eV. Taken together, the results in Table I and Figure 3, show that the VCA approach gives an adequate description of both the structural and electronic properties of LSMO without artificial variations in the rotation angle.

*Effect of strain on LSMO*

The effect of in-plane strain in the (111)- and (001)-plane is different, as earlier reported for LaAlO$_3$ [27], which has the same ground state as LSMO, $R\bar{3}c$. Relying on both VCA and supercell approaches, we find that in agreement with earlier results for (111)-strained LaAlO$_3$, tensile (111)-strain inflicted on LSMO from STO preserves the $R\bar{3}c$ symmetry with tilt pattern $a^-a^-a^-$. The effect of (111)-strain on the octahedral rotations is thus to increase $\alpha = \beta = \gamma$, and with the VCA approach 5.50° is obtained compared the bulk VCA value of 5.19°. On the other hand, (001)-strain from a STO substrate does not preserve the $a^-a^-a^-$ tilt pattern and an in-phase rotation is found experimentally [49, 50]. However, the exact tilt pattern and space group of (001)-oriented LSMO strained to STO is under debate. Boschker et al. [50] find an $a^-a^-c^+$ tilt pattern with $Pnma$ symmetry, in contrast, Vailionis et al. [49] find an $a^+a^-c^0$ tilt pattern with space group $Cmcm$ for LSMO on (001)-oriented STO. Our results with both the supercell approach and VCA agrees with Boschker et al. [50], where an in-phase octahedral rotation is preferred around the out-of-plane axis resulting in a $a^-a^-c^+$ tilt pattern with $Pnma$ symmetry. The octahedral rotation angles from the VCA approach are calculated at: $\alpha = \beta = 7.48°$ and $\gamma = 5.23°$. The DFT calculations point towards a non-zero out-of-plane rotation under tensile (001)-strain, which is in contrast to results for insulating oxides, e.g. LaAlO$_3$ [27, 31, 39]. However, it is in line with results for other metallic oxides, e.g. LaNiO$_3$, where the increased screening results in a non-zero out-of-plane rotation also under tensile strain [3, 32].

*Octahedral coupling across the (111)-interface*

Octahedral coupling from different phases in STO calculated with VCA and supercell to treat the Sr doping of LSMO is shown in Figures 4 and 5, respectively. The response is similar for both approaches. For $a^0a^0a^0$ locked STO, Figure 4 a and 5 a, the relaxation occurs almost exclusively in the STO layer closest to the interface (layer 0). Note that all the rotation angles $\alpha, \beta$ and $\gamma$ in this case respond similarly, as the six-fold inversion axis is preserved. That first LSMO layer is slightly affected by the $a^0a^0a^0$ locked STO, as shown in Figure 4a and 5a, implies that the STO is more susceptible to changes in octahedral rotations than LSMO, and that the octahedral rotation pattern would likely propagate further into STO if it had not been fixed in the calculations. On the other hand, for the $a^0a^0c^-$ locked STO (Figure 4b and 5b) the $\gamma$-rotation increases above the equilibrium value, denoted $\gamma_{strain}$, both in LSMO and STO before it reaches the equilibrium value in LSMO after approximately 7 layers ~ 14 Å. The same relaxation length is observed for the $\alpha$ and $\beta$ rotations, but they approach the equilibrium value, denoted $\alpha_{strain}$ from below. That the $\gamma$-angle increases above its equilibrium value before it is reduced for the $a^0a^0c^-$ locked STO (Figure 4 b and 5 b) is consistent with a lower energy cost when all rotation angles are similarly distorted, as opposed to having different relaxation lengths for the different rotation axes. There are two notable differences between the supercell approach and the VCA. (1) Discrete doping in the supercell causes a variation in the rotation angles in LSMO far away from the interface ($\sigma \neq 0$, Table I). (2) Since the equilibrium rotation angles are different in LSMO calculated with VCA or supercell approach (Table I), the equilibrium rotation angles differ also. Still, how the rotations couple across the interface is similar in terms of coupling length and which angles are increased or reduced. I.e., both the VCA and supercell approach can be used to model interface effects in STO/LSMO. However, the VCA approach is preferable when assessing properties such as relaxation lengths, as these properties then do not depend on the choice of discrete Sr positions. Based on this rationale, we will only consider interfaces calculated with the VCA approach in the following.

*Octahedral coupling across the (001)-interface*

The octahedral coupling across the (001)-interface between LSMO and STO is shown in Figure 6, while the corresponding coupling across the (111)-interface was presented in Figure 4. For the (001)-interface the in-plane ($\alpha$ and $\beta$) and out-of-plane ($\gamma$) rotations evolve differently, even for STO is locked in the $a^0a^0a^0$-rotation pattern (Figure 6 a), in contrast to (111)-interface coupling (Figure 4 a). As expected, the $\alpha = \beta$ rotations gradually increase towards the equilibrium values, $\alpha_{strain} = \beta_{strain}$, over 4 LSMO layers ~ 14 Å (Figure 6a). In contrast, the $\gamma$-rotation for the (001)-interface varies depending on the distance to the interface. We note that in the 3$^{rd}$ layer the rotation order is changed, the first two layers has out-of-phase rotation (−), as preferred in STO, and increased out-of-plane lattice parameter $c_{pc} \approx$ 3.84 Å. On the other hand, layers farther away from the interface has in-phase rotations (+) and lattice parameter of $c_{pc} \approx$ 3.82 Å. This is consistent with that the Pnma symmetry with tilt pattern $a^-a^-c^+$ is stabilized by tensile (001) strain [50]. A change of octahedral rotation pattern and increased lattice parameters close to the interface has also been documented for LSMO on STO(001) experimentally [51]. We note that there here are in-phase and out-of-phase octahedral rotations that meet, which for (111)-interfaces gives rise to a rotation mismatch and changed magnetic properties [30]. No such mismatch is expected in this case, as the change of rotation type occurs around the out-of-plane rotation axis.

When STO is locked in a tilt pattern equivalent to the tetragonal state stable below 105 K, there are two symmetry inequivalent orientations with respect to the (001)-interface, $a^0a^0c^-$ and $a^-b^0b^0$. The octahedral coupling for these two orientations are shown in Figure 6 b and c, respectively. For the $a^0a^0c^-$-locked STO, shown in figure 6 b, the coupling is almost identical to the coupling of $a^0a^0a^0$ locked STO with only minor differences in rotation amplitudes. This is in contrast to the case for $a^-b^0b^0$-locked STO, as shown in Figure 6 c. In this case, there is an additional difference between the in-plane rotations $\alpha$ and $\beta$. Here the $\alpha$-rotation, which is non-zero in STO, increases to the equilibrium value in LSMO, $\alpha_{strain}$ in the first LSMO layer, similar to what occurred for LSMO on $a^0a^0a^0$-locked STO(111) (Figure 4 a). On the other hand, $\beta$ which is zero in STO (Figure 6c), relaxes to its equilibrium value $\beta_{strain} = \alpha_{strain}$ over 4 unit cells ~ 14 Å, which is the same distance shown for LSMO on $a^0a^0a^0$- and $a^0a^0c^-$-locked STO (001) (Figure 6 a and b). The out-of-plane rotation angle $\gamma$ is slightly different for $a^-b^0b^0$-locked STO (Figure 6 c), compared to $a^0a^0a^0$- and $a^0a^0c^-$-locked STO (001) (Figure 6 a and b). For $a^-b^0b^0$-locked STO (Figure 6 c), the $\gamma$-rotation is lower in the first LSMO layers and never exceeds $\alpha_{strain}$. Also, in the 3$^{rd}$ LSMO layer, the $\gamma$-rotation is not reduced as much as for $a^-b^0b^0$-locked STO compared to $a^0a^0a^0$- and $a^0a^0c^-$-locked STO (001). These significant differences between $a^0a^0c^-$ and $a^-b^0b^0$ locked STO shows that for (001)-interfaces, the in-plane rotations couple more strongly into the thin film than out-of-plane rotations. We note that this is consistent with tensile (001)-strain favoring in-plane rotations in the strained material.

*Octahedral coupling length*

As shown in the previous sections, there are both similarities and differences between the octahedral coupling in the (111)- and (001)-plane. For the (111)-interface a coupling length of 7 layers, corresponding to ~ 14 Å was found, while for the (001)-interface the coupling is longer for in-phase out-of-plane rotations ($\gamma$) than out-of-phase in-plane rotations ($\alpha$ and $\beta$). Hence, the total octahedral coupling length is longer for (001)-oriented LSMO compared to the (111)-orientation, as (111) strain does not give rise to in-phase rotations. However, if one only considers the out-of-phase rotations, the coupling length for the (001)-interface is 4 unit cells, which corresponds to a length of ~ 14 Å, as also found for the (111)-interface. The similar coupling length for (001) and (111) interface facets indicates that the coupling is a steric effect, in agreement with Aso *et al* [11].

*Effect on spin densities*

In order to elucidate how the rotation pattern of the STO can affect the functional properties of an epitaxial thin film, the calculated spin density difference in LSMO, depending on if STO is locked in a tilt pattern of $a^0a^0c^-$ and a $a^0a^0a^0$, is presented in Figure 7. For the (111)-interface all three variations of the $a^0a^0c^-$ tilt pattern are symmetry equivalent, while for the (001)-interface the $a^0a^0c^-$ and $a^-b^0b^0$ differ. The spin density difference shows spatial variations in the spin polarized electron density, even though the total at each Mn site is almost unaffected ($\Delta M \leq 0.013 \ \mu_B$) by the octahedral rotations in STO. The almost constant moment per Mn can be explained from the large bandwidth of the half-metallic state of LSMO, as shown in Figure 3 [10].

In Figure 7 a, it is shown that effect of the octahedral tilt pattern in STO below the 105 K structural transition has a large effect on the spin density in LSMO on (111)-oriented STO. For this particular condensed $a^0a^0c^-$ tilt pattern, the change is a net shift of the spin density along the in-plane $[1\bar{1}0]$-direction, as shown in Figure 7 c) by the blue and yellow lobes. The difference is largest at the interface, and it is reduced further into the LSMO (Figure 7a), consistent with the previous finding that the octahedral rotations relaxes towards the same value far away from the interface (Figure 4). The shift along the $[1\bar{1}0]$-direction can be rationalized from the $[1\bar{1}0]$-axis being orthogonal to the locked STO rotation axis around [001].

In Figure 7 b), d) and f) it is shown that for the (001)-interface, there is a large difference between the two symmetry inequivalent versions of STO rotations. When the STO is locked in a state with out-of-plane rotations ($a^0a^0c^-$) the spin density difference is almost zero (Figure 7 b). There is only a weak signal in the spin density difference in the 4$^{th}$ Mn layer, with a difference along the in-plane $[\bar{1}10]$-direction as shown in Figure 7 e), orthogonal to the STO rotation axis [100]. This small difference in spin density can be related to the relatively small change in octahedral rotations when the STO is locked an $a^0a^0a^0$ tilt pattern, compared to $a^0a^0c^-$ (Figure 6 a and b). On the other hand, when the STO layer is locked with an in-plane tilt pattern, e.g. $a^-b^0b^0$, there is a substantial spin density difference as shown in Figure 7 b) and e). Similar to the case for the (111)-interface (Figure 7 a), the spin density difference for STO(001) locked in $a^-b^0b^0$ tilt pattern (Figure 7c) is largest at the interface and is reduced further into the LSMO as the difference between the tilt patterns is reduced (Figure 6 a and c). Furthermore, for the $a^-b^0b^0$ locked STO(001), the spin density difference is now along [011] as shown in Figure 7 e), orthogonal to the STO rotation axis [100]. Hence, in contrast to the (111)-oriented interface, (001)-interfaces can give rise to a spin density difference with an out-of-plane component.

Comparing the absolute values of the spin density shifts we find that the (111)-interface has an absolute difference of 0.078 q/Å$^2$, while the absolute value the (001)-interface is 0.022 q/Å$^2$ and 0.092 q/Å$^2$ depending on whether the rotations in the STO is out-of-plane or in-plane, respectively. I.e. the absolute shifts for (111) and (001) in-plane rotated STO are similar and considerably larger than (001) with the rotation axis out-of-plane.

The difference in spin density can affect the magnetic properties such as the easy axis and magnetic moment. For thin films, the magnetic easy axis is typically in the film plane [52], and LSMO on cubic (001)-oriented STO has biaxial anisotropy, with easy axes along the ⟨110⟩-family of in-plane directions [53]. Furthermore, it has been shown that when LSMO on (001)-STO is cooled below the tetragonal transition temperature, one observes an increase in the coercive field which can be controlled by the cooling history [14]. For LSMO on (111)-oriented cubic STO on the other hand, six different magnetic easy axes along both the ⟨1$\bar{1}$0⟩- and ⟨11$\bar{2}$⟩-families of in-plane directions are found at room temperature experimentally [54]. For a given $a^0a^0c^-$ condensation, corresponding to Figure 7 a and c, there is an effective spin accumulation along the [1$\bar{1}$0] direction below 105K, as shown in Figure 8. If all three $a^0a^0c^-$ rotational variances condense, we predict that the tetragonal transition in STO should lift the degeneracy between the ⟨1$\bar{1}$0⟩- and ⟨11$\bar{2}$⟩-family of in-plane directions and result in triaxial anisotropy. These changes in coercive field can be rationalized in terms of different domains, which have different effects on the spin density.

**Conclusions**

The different symmetry of the (111)- and (001)-interfaces lead to significantly different octahedral coupling between LSMO and STO, where the (111)-interface shows a gradual transition towards the equilibrium rotation amplitude, while the (001)-oriented LSMO has an additional change from an in-phase (+) to an out-of-phase (−) rotation. However, the coupling length of out-of-phase rotations is the same, ∼ 14 Å, pointing towards octahedral coupling being a steric effect. Furthermore, for the (111)-interface, the $a^0a^0c^-$ tilt pattern of STO have three symmetry equivalent variances, while for the (001)-interface, the out-of-phase rotation in STO can be either in-plane ($b^-a^0a^0$ or $a^0b^-a^0$) or out-of-plane ($a^0a^0c^-$). For (001)-oriented LSMO the effect of the $a^0a^0c^-$ rotation in STO is similar to that of $a^0a^0a^0$ STO, while the $b^-a^0a^0$ STO gives different coupling lengths for the $\alpha$- and $\beta$-rotaions, and reduced the impact on the $\gamma$-rotation. In addition, the octahedral coupling affects the spin density, where the difference is perpendicular to the STO-rotation axis. These results demonstrate that one can rely on geometrical lattice engineering and steric effects to control octahedral coupling and the directional dependence of functional properties.

**Acknowledgements**

The Norwegian Metacenter for Computational Science is acknowledged for providing computational resources, Uninett Sigma 2, Project No. NN9301K. TT acknowledges funding through the Research Council of Norway grant No.231290.

**ORCID**

Magnus Moreau http://orcid.org/0000-0003-0046-7375

Sverre M. Selbach http://orcid.org/0000-0001-5838-8632

Thomas Tybell http://orcid.org/0000-0003-0787-8476

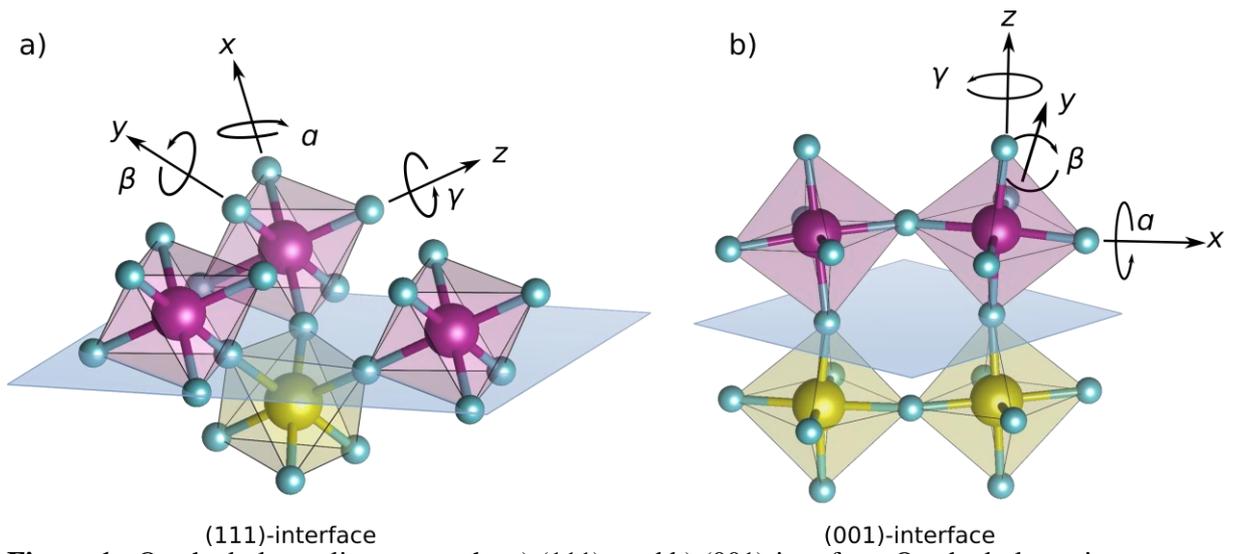

**Figure 1:** Octahedral coupling across the a) (111)- and b) (001)-interface. Octahedral rotation angles α, β and γ the pseudocubic x-, y- and z-directions, respectively, are illustrated.

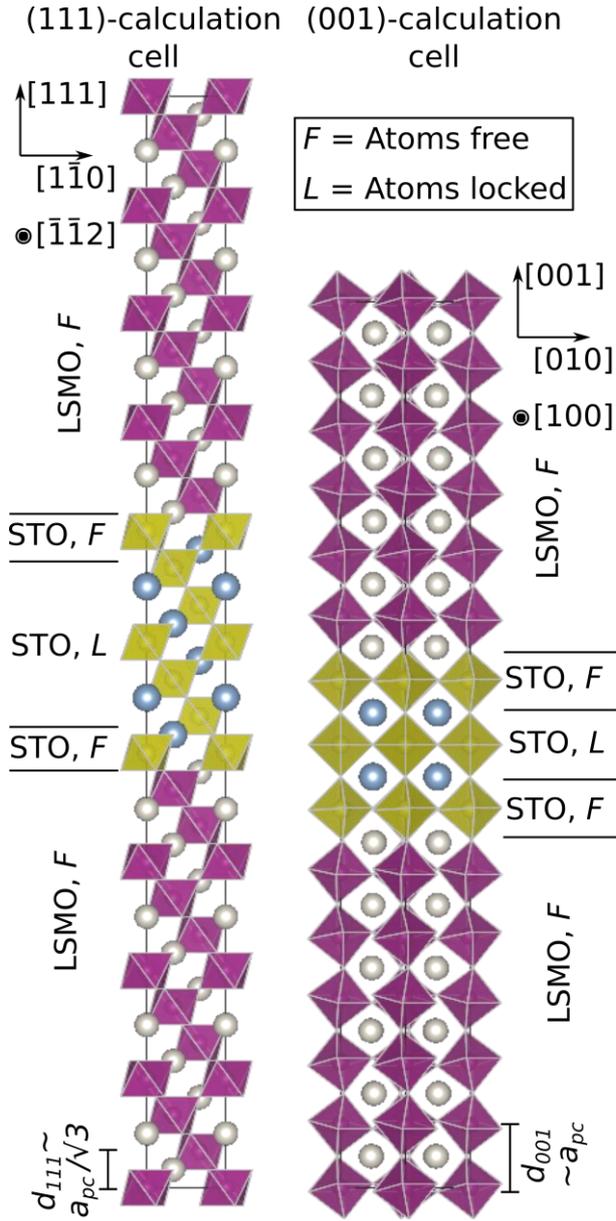

**Figure 2:** $\sqrt{2} \times \sqrt{2} \times 10\sqrt{3}$ (left) and $\sqrt{2} \times \sqrt{2} \times 14$ (right) supercells used for calculating octahedral coupling across (111)- and (001)-interfaces, respectively. Ionic positions in the STO layers away from the interface were locked in relaxed bulk positions, with tilt pattern $a^0a^0a^0$ (cubic) or $a^0a^0c^-$ (tetragonal). For the (001)-interface, the rotation axis can be either out-of-plane, i.e. $a^0a^0c^-$, or in-plane, i.e. $a^-b^0b^0$, which is equivalent to $a^0b^-a^0$. For (111)-strain these variations are symmetry equivalent.

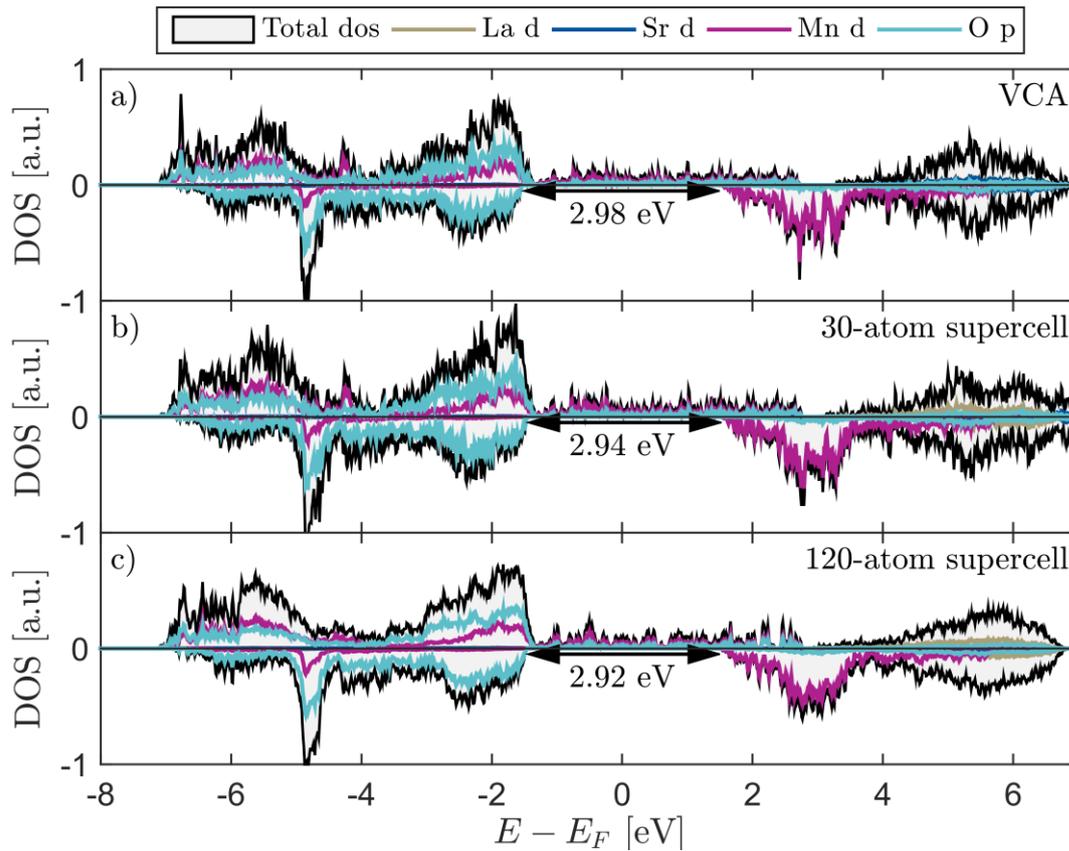

**Figure 3:** Total and orbital projected Density of States (DOS) comparing the different approaches of modelling the Sr doping in LSMO. a) VCA, b) 30 atom supercell, c) 120 atom supercell. The black arrows indicate the half-metallic bandwidth, the energy region with only spin up states.

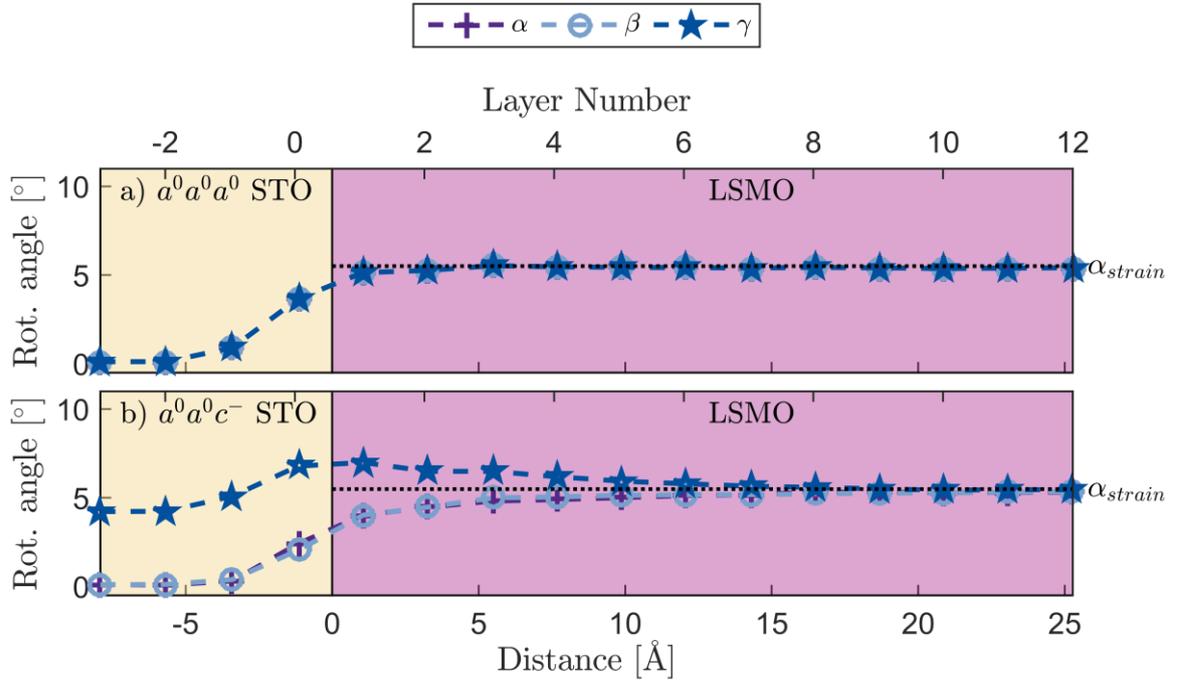

**Figure 4:** Octahedral rotational coupling across the (111)-interface between LSMO and STO, calculated with the VCA approach, where STO is fixed to be a) cubic ($a^0a^0a^0$) and b) tetragonal ($a^0a^0c^-$). The STO layer closest to the interface (layer 0) was allowed to relax. Rotation angles $\alpha, \beta$ and $\gamma$ are defined in Figure 1 a). $\alpha_{strain}$ is the rotation angle calculated for LSMO when strained to an (111)-oriented STO substrate.

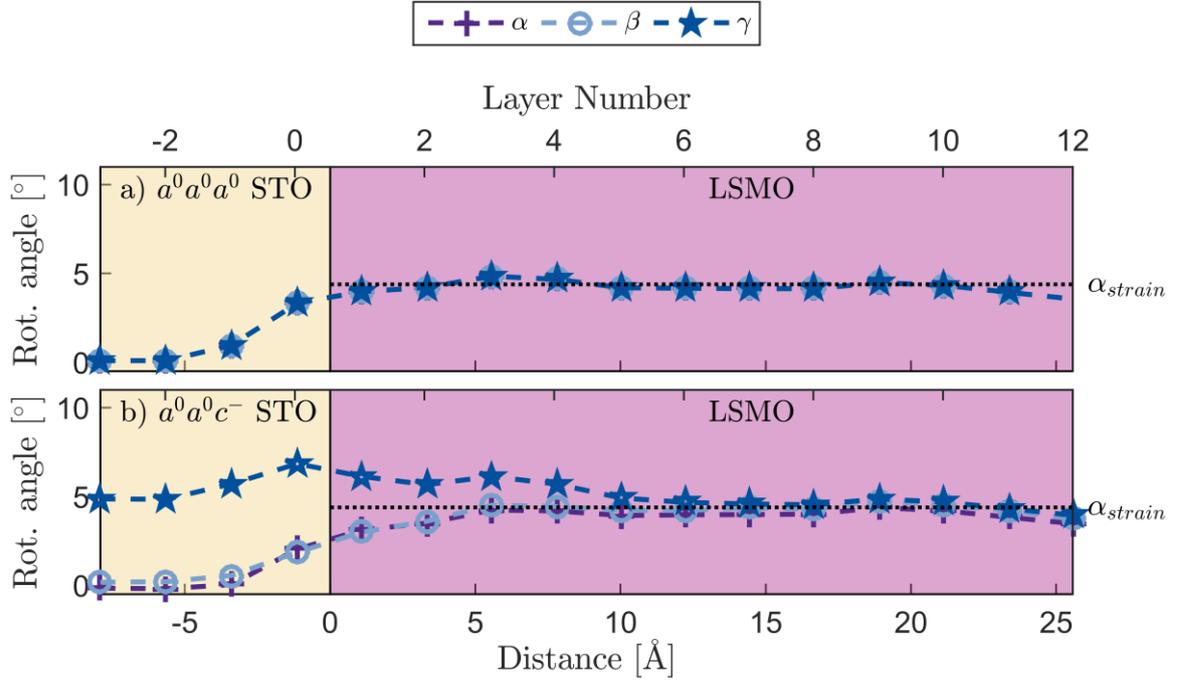

**Figure 5:** Octahedral rotational coupling across the (111)-interface between LSMO and STO, calculated with an in-plane $\sqrt{2}\times\sqrt{2}$ supercell approach, where STO is fixed to be a) cubic ($a^0a^0a^0$) and b) tetragonal ($a^0a^0c^-$). The STO layer closest to the interface (layer 0) was allowed to relax. The rotation angles $\alpha, \beta$ and $\gamma$ are defined in Figure 1 a). $\alpha_{strain}$ is the rotation angle calculated for LSMO when strained to an (111)-oriented STO substrate.

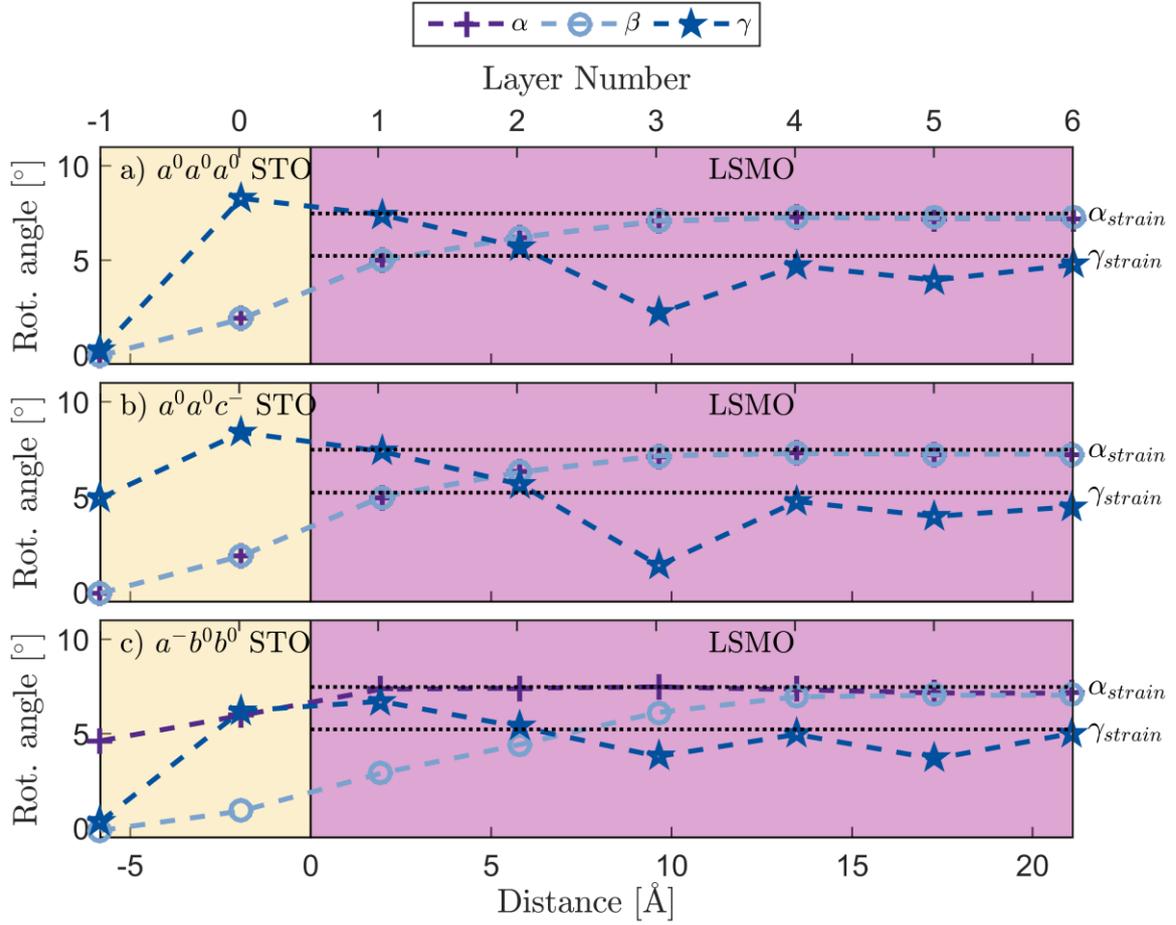

**Figure 6:** Octahedral rotational coupling across the (001)-interface between LSMO and STO calculated with the VCA apprach with STO fixed to be a) cubic ($a^0a^0a^0$), and b-c) tetragonal with tilt patterns b) $a^0a^0c^-$ and c) $a^-b^0b^0$. The octahedral rotation axis is out-of-plane in b) and in-plane in c). The STO layer closest to the interface (layer 0) was allowed to relax. The rotation angles $\alpha, \beta$ and $\gamma$ are defined in Figure 1 b). $\alpha_{strain}$ and $\gamma_{strain}$ are the rotation angles calculated for LSMO when strained to an (001)-oriented STO substrate.

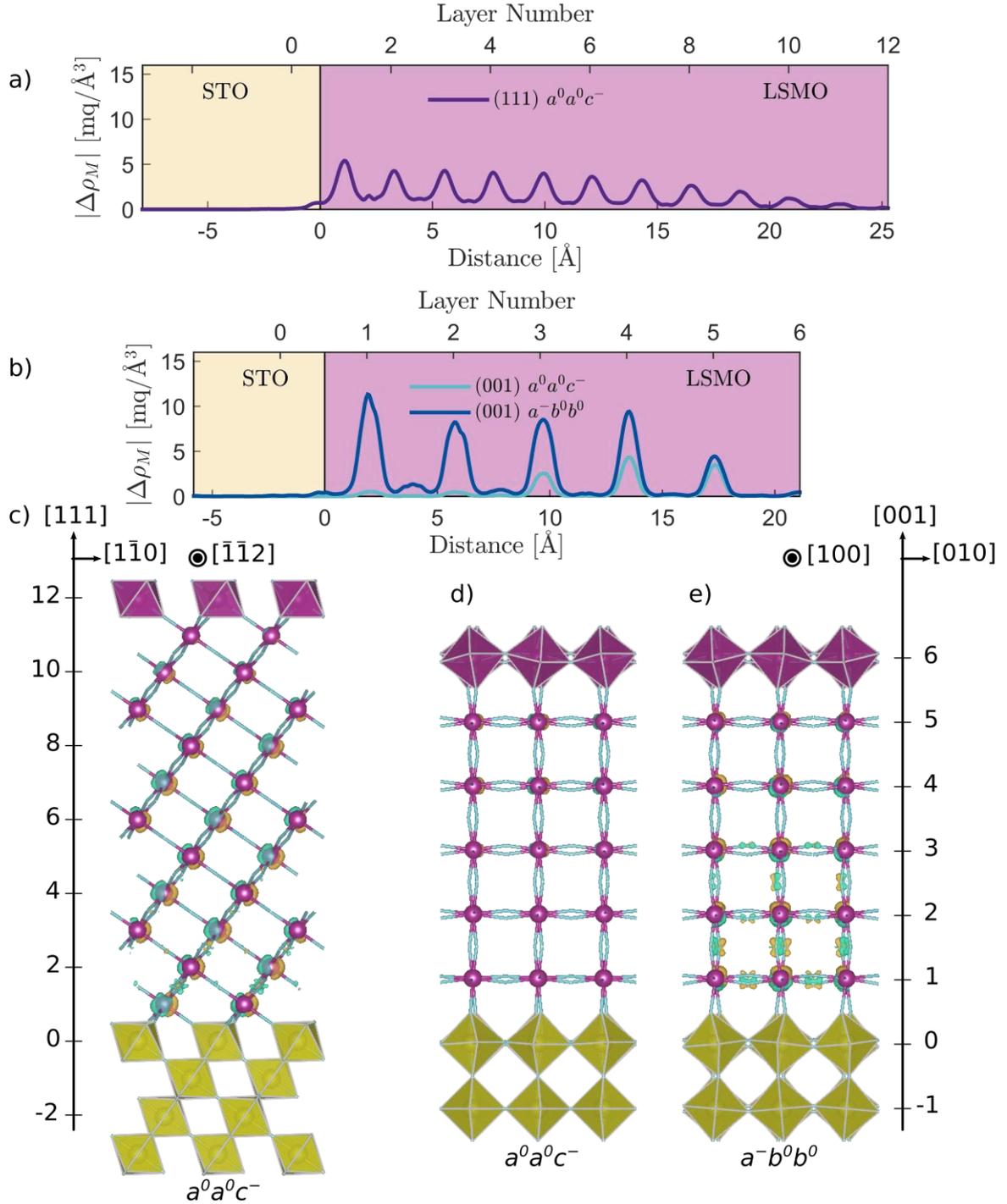

**Figure 7:** Calculated spin density difference, $\Delta\rho_M$, between STO locked in cubic structure with $a^0a^0a^0$-tilt pattern and in two different tetragonal tilt patterns. The absolute value of $\Delta\rho_M$ as a function of distance from the a) (111) and b) (001) interface. c-e) Isosurface plots of the spin density difference superimposed on the atomic structure for c) $a^0a^0c^-$ tilt pattern across a (111)-interface, d) $a^0a^0c^-$ tilt pattern across a (001)-interface and e) $b^-a^0a^0$ tilt pattern across an (001)-interface. Positive and negative is colored yellow and blue, respectively. The isosurface level is set to 5 mq/Å³.

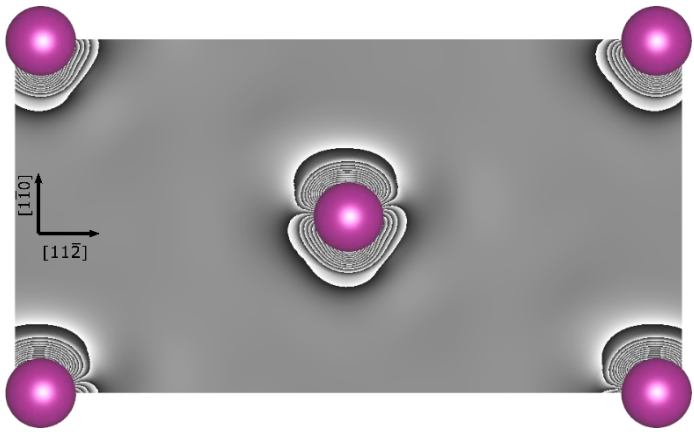

**Figure 8:** Contour plot of the spin density difference in the (111)-plane closest to the interface showing a clear anisotropy between the [1$\bar{1}$0]- and [11$\bar{2}$]-directions below 105 K. Purple spheres represent Mn atoms.

**TABLE I**: Comparison of bulk parameters calculated for LSMO with VCA, $\sqrt{2} \times \sqrt{2} \times 2\sqrt{3}$ (30 atoms) and $2\sqrt{2} \times 2\sqrt{2} \times 2\sqrt{3}$ (120 atoms) supercells of the aristotype perovskite and experimental values. $a, b$ and $c$ represent the lattice parameters of LSMO in the $R\bar{3}c$ space group in the hexagonal setting, $M$ is the saturation magnetism, $\alpha, \beta$ and $\gamma$ represent the mean octahedral rotations around the pseudocubic axes, $\sigma$ represents the standard deviation of the octahedral rotations, $E_{AFM} - E_{FM}$ is the exchange energy, defined as the difference between a G-type antiferromagnetic and a ferromagnetic spin ordering. Experimental data from [48].

| Parameter | VCA | 30-atom supercell | 120-atom supercell | Experimental |
|---|---|---|---|---|
| $a = b$ [Å] | 5.500 | 5.510 | 5.509 | 5.506 |
| $c$ [Å] | 13.214 | 13.322 | 13.351 | 13.356 |
| $c/\sqrt{6}a$ | 0.981 | 0.987 | 0.989 | 0.990 |
| $M$ [$\mu_B$/Mn] | 3.67 | 3.67 | 3.67 | 3.67 |
| $\alpha = \beta = \gamma$ [°] | 5.19 | 4.45 | 4.25 | 4.48 |
| $\sigma$ [°] | 0 | 0.187 | 0.310 | |
| $E_{AFM} - E_{FM}$ [eV/f.u.] | 0.423 | 0.386 | 0.221 | |
| CPU cost [hours/16 cores] | 2.7 | 8.4 | 177 | |